# Equivalent el. circuit of magnetosphere-ionosphere-atmosphere interaction

## P.A. Sedykh

Siberian Branch of Russian Academy of Sciences, Institute of Solar-Terrestrial Physics, Lermontov str., 126 a, p/o box 291, Irkutsk, 664033, Russia

E-mail: pvlsd@iszf.irk.ru

## Abstract

The aim of this study is to investigate the magnetospheric disturbances effects on complicated nonlinear system of atmospheric processes. During substorms and storms, the ionosphere was subjected to rather a significant Joule heating, and the power of precipitating energetic particles was also great. Nevertheless, there were no abnormal variations of meteoparameters in the lower atmosphere. If there is a mechanism for the powerful magnetospheric disturbance effect on meteorological processes in the atmosphere, it supposes a more complicated series of many intermediates, and is not associated directly with the energy that arrives into the ionosphere during storms. I discuss the problem of the effect of the solar wind electric field sharp increase via the global electric circuit during magnetospheric disturbances on the cloud layer formation.

## 1. Introduction

The book (Herman and Goldberg, 1978) is particularly welcome at a time at which many astronomers, space scientist, geophysicist, and meteorologists are entering the field of Sun-weather/climate investigation. This book provides an excellent opportunity for a scientist considering this new field to get an overall view of the present status of the subject in its many disciplinary aspects. The existing correlations provide a strong suggestion that some physical mechanism exists linking the variable Sun and the weather and climate, but the details of such a mechanism or mechanisms are quite unknown.

Statistical correlations were found between geomagnetic activity, atmospheric pressure and temperature (Bucha, 2009), (Bucha, 2012), (Palamara, 2004). Authors of (Haigh et al., 2005) suggested that the observed climate response to solar variability is brought about by a dynamical response in the troposphere to heating predominantly in the stratosphere.

It is known that a tropical cyclogenesis may be "a mechanism for effective discharge of the surplus heat in the atmosphere under the conditions when the routine mechanism effect becomes insufficient" (Sharkov, 2005). Between the solar-terrestrial disturbance parameters, on the one hand, and the cyclogenesis characteristics, on the other, various researchers endeavor to trace hard-to-detect statistical communications associations. The revealed coincidence between the time of origin and evolution of the 23-24 Aug 2005 Hurricane Katrina with the powerful geomagnetic storm main phase (Ivanov, 2006) also boosted the research in this trend.

The issue of the reality and the physical mechanism for solar-terrestrial couplings has rather a long history. Many geophysicists were most decisive in rejecting the idea about a solar activity effect on the lower atmosphere condition as absolutely unacceptable. And, first of all, the matter was that the power of atmospheric processes exceeds the solar-wind input energy flux into the near-Earth space enormously. Due to this, it seems most unlikely that solar activity could significantly affect the lower atmosphere condition. However, the research done over the last years allowed us to find a clue to overcoming this inconsistency. The main objection to a possibility of the solar activity effective influence on the condition of the lower atmosphere and weather, based on insufficient power of the solar wind, appears quite surmountable; see e.g. (Pudovkin, 1996). Also, like the computations in (Pudovkin, 1996) show, the energy necessary to create the atmospheric optical screen (shield) is incomparably lower than the amplitude of the variation in the screen-induced solar energy flux arriving in the lower atmosphere.

According to (Pudovkin, 1996), a noticeable variation in the chemical composition and contents of small components, as well as in the atmosphere transparency, is caused by variations in the atmospheric ionizing radiation flux observed during geomagnetospheric disturbances. The main types of such variations are (Pudovkin, 1996): 1) the galactic cosmic ray intensity short-term depressions observed during geomagnetic disturbances (Forbush decreases) caused by dispersion of energetic charged particles by the magnetic fields transported from the solar atmosphere by the solar wind high-velocity streams; 2) solar cosmic ray flux bursts caused by solar flares.

The link between the enhanced solar wind (geomagnetic) forcing, the North Atlantic Oscillation (NAO) and changes in the troposphere was suggested in (Bucha et al., 1998) showing that the strengthening of thermospheric winds generates vertical downward winds in the aurora (Crowley et al., 1989).

In (Mansilla, 2011) author examined the possible connection between atmospheric parameters measured at low and middle altitudes and geomagnetic storms occurred in 2000 and 2003. The results presented in (Mansilla, 2011) may show evidence to support that atmospheric parameters at heights of the troposphere and lower stratosphere could be possibly related to geomagnetic storms.

According to (Troshichev, Janzhura, 2004), the interplanetary electric field influence is realized through acceleration of the air masses, descending into the lower atmosphere from the troposphere, and formation of cloudiness above the Antarctic Ridge, where the descending air masses enter the surface layer. The cloudiness results in the sudden warming in the surface atmosphere, because the cloud layer efficiently backscatters the long wavelength radiation going from ice sheet, but does not affect the process of adiabatic warming of the descending air masses. Influence of the interplanetary electric field on cloudiness has been revealed for epochs of the solar activity minimum, when Forbush decreases effect is absent. The acceleration of the descending air masses is followed by a sharp increase of the atmospheric pressure in the near-pole region, which gives rise to the katabatic wind strengthening above the entire Antarctica. As a result, the circumpolar vortex around the periphery of the Antarctic continent decays and the surface easterlies, typical of the coast stations during the winter season, are replaced by southerlies. It is suggested that the resulting invasion of the cold air masses into the Southern ocean leads to destruction the regular relationships between the sea level pressure fluctuations in the Southeast Pacific High and the North Australian –Indonesian Low, since development the El-Nino event strongly follows anomalous atmospheric processes in the winter Antarctica.

A magnetospheric storm is a 1-3 day long phenomenon spanning all the magnetosphere regions, and it features sharp depressions in the magnetic field. During storms and substorms, the ionosphere undergoes rather significant Joule heating with a great power of precipitating energetic particles. Huge energy increases the ionosphere temperature, causes large-scale ion drifts, and neutral winds. Nevertheless, there were no abnormal variations in the lower atmosphere meteoparameters (Sedykh, Lobycheva, 2013). The variations in physical characteristics of hurricanes (wind velocity, temperature and pressure distribution) were not associated with high magnetic activity both in the high-latitude regions, and in the mid-latitude ones. We have used a lot of data for the analysis; some examples are illustrated in figures (Sedykh, Lobycheva, 2013). There is no principal difference in plots for the periods of magnetospheric disturbances in comparison to nonstorm days. The researched high-latitude region ($60^o$-$90^o$) has been divided into six zones with characteristic types of temperature changes (according to the paper (Panin et al., 2008)) to study changes of geomagnetic activity and meteorological parameters: Pacific, Canadian, Atlantic, European, Siberian and Far East. The math statistical analysis has shown low correlation dependence between total power of Joule heating of the ionosphere (and power of particle precipitations) and change of meteorological parameters (temperature, geopotential heights) in the allocated six zones for all above mentioned selected periods of magnetospheric disturbances. The Joule heating during extra powerful magnetospheric disturbances is effective in the ionosphere but it does not play key roles in the lower atmosphere.

The objective of this paper is to investigate possible effects of geomagnetic activity on meteorological processes in the atmosphere.

## 2. The magnetosphere-ionosphere-atmosphere interaction

Cloud layers play an important role in Earth's radiation balance (Tinsley et al., 1989), (Tinsley et al., 1991), (Tinsley et al., 1993), affecting the amount of heat from the Sun that reaches the surface and the heat radiated back from the surface that escapes out into space. One should pay special attention to the effect of the solar wind electric field sharp increase (via the global electric circuit during magnetospheric disturbances) on the cloud layer formation. It is necessary to test the assumptions that this layer may function as a screen decelerating radiative cooling of the air located on the Central Antarctic ice dome (as a result, there would be warming in the ground atmospheric layer, and cooling above the cloud layer (Troshichev, Janzhura, 2004)). Authors in the paper (Troshichev, Janzhura, 2004) suggested that the interplanetary electric field influences the katabatic system of atmospheric circulation (typical of the winter in the Antarctic), via the global electric circuit affecting clouds and hence the radiation dynamics of the troposphere.

Let us address the problem of the extraneous electric field penetration into the Earth's magnetosphere. More than five decades ago, Dungey (Dungey, 1961) suggested the following model for the magnetospheric electric field generation. The solar wind magnetic field partially permeates inside the magnetosphere, therefore, the polar cap field lines leave for the solar wind. In the solar wind, there is the electric field **E=-(1/c)[VB],** where V is the solar wind velocity, B is the interplanetary magnetic field. Because the conductance along field lines is very high, the electrical potential associated with the field in the solar wind is transported into the polar cap ionosphere. In the polar cap, a Sun-away convection originates, and, for that convection to be closed, reverse motions on closed field lines of the inner geomagnetosphere are necessary. Two-vortex convection was assumed to be obtained. Through multiplying the solar wind electric field value by the geomagnetosphere size towards dawn-dusk being about 25-30 Re (Re~6371 km), and through assuming the solar wind velocity as ~450 km/s, we will obtain the potential difference between the dawn and the dusk sides of the magnetosphere $\Delta\psi\approx$-80Bz, (kV) which is manifold less than the well-known experimental formula (Doyle, Burke, 1983) $\Delta\psi$(kV)=-11Bz (nT) provides.

The penetration of the electric field and the current into the geomagnetosphere is a two-stage process, and may be presented as follows. Let an electric current component towards the magnetosphere appear at instant T. A potential value will be established at the magnetopause segment. In the thin near-side layer of the thickness $d \sim 2\pi c/\omega_{pp}$ (where $\omega_{pp}$ is proton plasma frequency, c is the speed of light) the charge division process will start, and the displacement current $\mathbf{j^*} = (\varepsilon/4\pi)\times\partial\mathbf{E}/\partial t$ will appear. Also, there will appear Ampere force $\mathbf{F} = [\mathbf{j}^* \times \mathbf{B}]/c$ that will start accelerating plasma. The only force that withstands the Ampere one is the inertia force. Under the conditions of a homogeneous medium, the inertia force is $\rho \partial \mathbf{v}/\partial t$:

$$\rho \partial \mathbf{v}/\partial t = [\mathbf{j}^* \times \mathbf{B}]/c = (\varepsilon/4\pi c) \times [\partial \mathbf{E}/\partial t \times \mathbf{B}].$$

Taking into account that $\varepsilon = c^2/V^2_A$, where $V_A$ is the Alfven velocity, upon integrating we will have: $\mathbf{v} = c[\mathbf{E} \times \mathbf{B}]/B^2$ that is the classic equation for the electric drift velocity (it is important for us to express the dynamic process in this case).

When the plasma is accelerated in the layer d up to the $\mathbf{V} \times \mathbf{B}$ drift velocity (and it happens during the gyroperiod), then there will be no field in the plasma coordinate system, and it appears at the boundary between the moving and stable plasmas in a stable coordinate system. The boundary moving velocity separating the moving plasma from the stable one will be, consequently, $V_\phi \sim d\omega_B/2\pi$, where $\omega_B$ is the proton gyrofrequency. Taking the values d and $\omega_B$ into the equation for the phase velocity, we see that it is the Alfven velocity like we expected. Thus, the external electric field penetrates into the magnetosphere without any limitations of the Alfven-wave type, and the electric current is only in a form of the displacement current. The electric current flows through the system under consideration only when there is a transitive process. In the stationary regime, there is no electric current. If the magnetic field is inhomogeneous according to an axis X, then the gas pressure gradient will be originated independently due to the flow nonuniformity. The electric field establishment time in the system here is $\tau_E = L/V_\phi$, and the current establishment time is $\tau_I = L^*/V_c$, where L is the system size, $V_\phi$ is the phase velocity for the electromagnetic signal propagation across the system, $L^* = (B/\nabla B)$, $V_c$ is the plasma convection velocity. An approximate estimate applied to the magnetosphere gives the time of the electric field establishment to be hundreds of seconds, and the electric current establishment time to be about an hour. Thus, the electric current penetration into plasma is a two-stage process. Initially, the polarization field that penetrates into plasma "layer by layer" is produced. Or, to be more exact, the momentum corresponding to this field penetrates into plasma. Here, if the system is inhomogeneous, the flow can redistribute pressure so that an electric current arises in plasma because of the appearance of gradients. This electric current is necessary to maintain plasma convection in the magnetosphere (Ponomarev, Sedykh, 2006(b)), (Sedykh, 2011).

The ionosphere is the ohmic environment where the electric field and current are related by the Ohm's law. Since the mean free path in the magnetosphere during pair collisions with a Coulomb interaction considerably exceeds the extents of the magnetosphere, it is customary to assume that the magnetospheric plasma is collisionless. A direct relation between the electric field and current is absent in the magnetosphere. Since the geomagnetic field lines are equipotential, the currents in the ionosphere and magnetosphere depend on the magnetospheric electric field and gas pressure distribution, respectively. If the ionospheric current was a purely Hall current, this would not be a dangerous phenomenon since the Hall current is nondivergent and does not deliver a work. In reality, the ionospheric current is combined and always includes the Pedersen component, and the ionosphere is a real energy consumer. Combined actions of some processes in the geomagnetosphere result in the formation of a spatial distribution of gas pressure, *i.e.*, bulk currents in the magnetosphere. Divergence of these bulk currents produces the spatial distribution of field-aligned currents. Thus, the problem of formation of a spatial distribution of plasma pressure in the magnetosphere is still very important. It is especially interesting that the necessary conception has long been known. This is the conception (Ponomarev, Sedykh, 2006 (a))**,** who reasonably indicated that the combination of pitch-angle diffusion and convection as a basis magnetospheric process is of importance. A combined action of plasma convection and pitch-angle diffusion of electrons and protons leads to the formation of plasma pressure distribution in the magnetosphere on the night side, and, as it is known, steady electric bulk currents are connected to distribution of gas pressure. The divergence of these bulk currents brings about a spatial distribution of field-aligned currents, i.e. magnetospheric sources of ionospheric currents. The projection (mapping) of the plasma pressure relief onto the ionosphere corresponds to the form and position of the auroral oval. This projection, like the real oval, executes a motion with a change of the convection electric field, and expands with an enhancement of the field.

The atmospheric process power incomparably exceeds energy flux from the solar wind into the geomagnetosphere, and the power of extremely strong magnetospheric disturbances. The energy flux from the magnetosphere into the atmosphere during the strong storm was about $1.5 \times 10^{19}$ (erg/s) $\times 24 \times 3600 = 1.2 \times 10^{24}$ ergs/day, which is by 2-3 orders of magnitude less than the atmospheric process power whose values are in (Pudovkin, Babushkina, 1992).

In the frame of the global electric circuit concept, thunderstorms act as a "meteorological" generator and create a potential drop of $U_{int} \approx 270$ kV between the ionosphere and the Earth's surface. It is considered that this potential is identical at all points of the ionospheric shell. In the high-latitude ionosphere, $U_{int}$ is superposed by the potential from the magnetospheric source ($U_{ext}$). Distribution of $U_{ext}$ corresponds to the ionospheric plasma convection, which is directly related to plasma convection in the geomagnetosphere. Therefore, how the electric field is transferred from the solar wind to the geomagnetosphere, and magnetospheric plasma convection generation, are very important issues.

Indeed, there is no simple global electric circuit via which a sharp increase in the solar wind electric field during magnetospheric disturbances would be possible. The solar wind electric field penetration process is complex and non-linear. Field-aligned currents connect the magnetosphere and the ionosphere in the united electric circuit. The plasma convection generation in the geomagnetosphere is associated with the processes at the bow shock front (Ponomarev, Sedykh, 2006 (b)). The combined action of plasma convection and pitch-angle diffusion of electrons and protons leads to the formation of plasma pressure distribution in the magnetosphere. As known, steady electric bulk currents are associated with the gas pressure distribution. The divergence of these bulk currents causes a spatial distribution of field-aligned currents, i.e., magnetospheric sources of the ionospheric electric current systems (Sedykh, Ponomarev, 2012; Sedykh, 2017). The atmospheric conductivity sharply declines between the polar ionosphere and the layer at h~10 km.

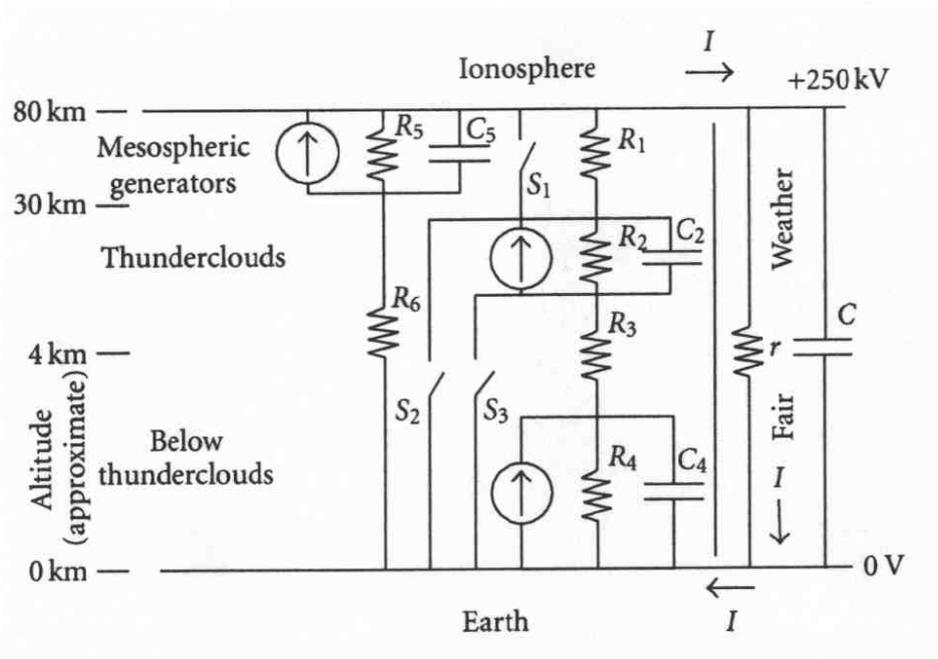

**Figure 1.** Diagram showing a schematic equivalent circuit for atmospheric electric circuits, credit (Rycroft, 2006).

A realistic model of equivalent circuit with capacitors, resistors, and switches is presented and is shown in figure 1 in (Rycroft, 2006). The electrodynamic coupling between the Earth's atmosphere and the ionosphere is very complex and may be described by the global electric circuit (Rycroft, 2006). The diagram in (Rycroft, 2006) showing a schematic equivalent circuit for global electric circuits.

## 3. Discussion and conclusions

The global atmospheric electric circuit is connected through a high-altitude ionosphere, and magnetospheric disturbances can effect on the stationary and changes of an atmospheric electric field. Process of electric field penetration from the solar wind is complicated; this phenomenon is nonlinear. Plasma convection generation in the geomagnetosphere is associated with processes at the bow shock front.

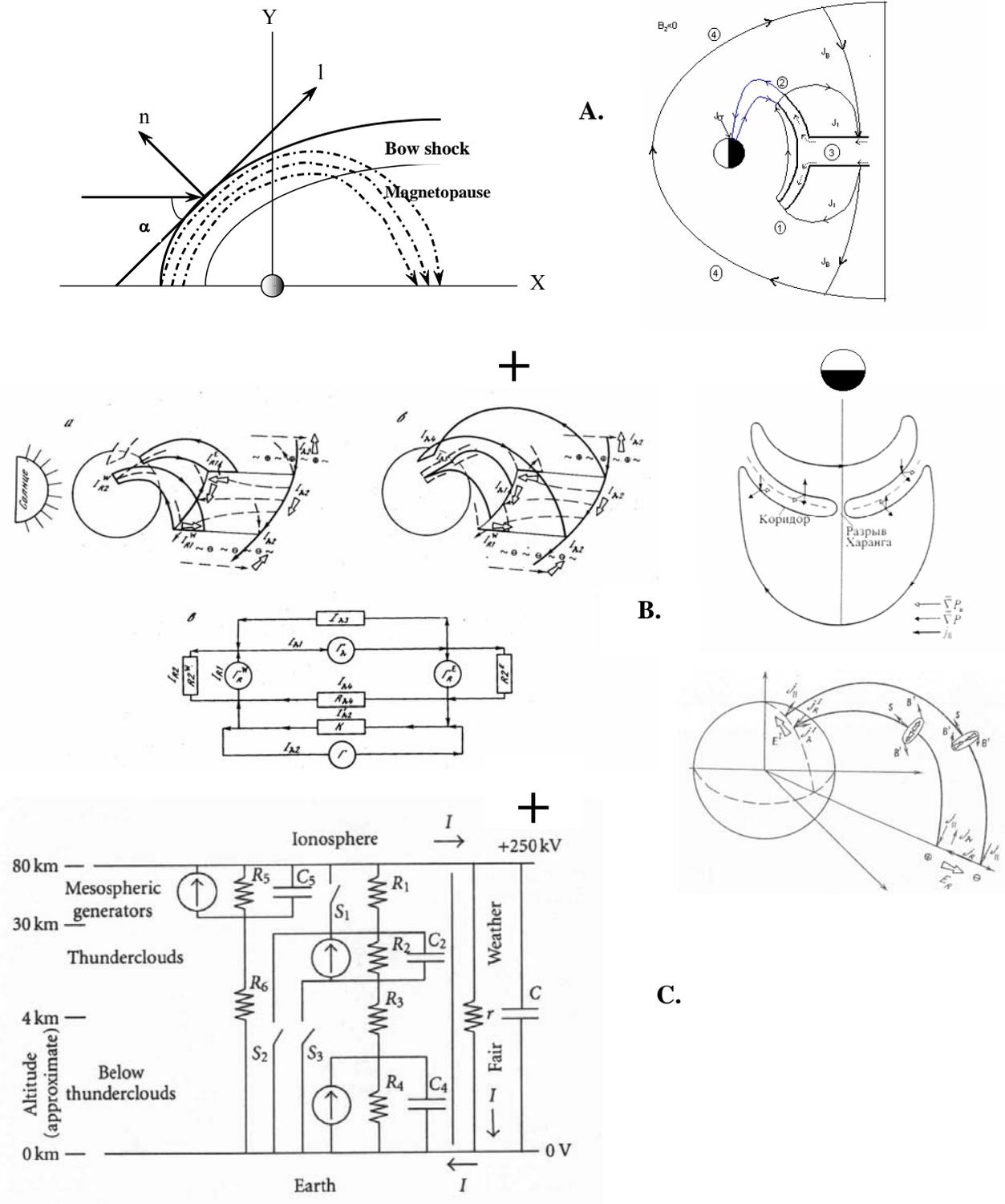

**Figure 2.** The scheme of locations of functional blocks in the magnetosphere(A); (B) schematic spatial location of the magnetospheric-ionospheric currents (Ponomarev, Sedykh, 2006 (a)).; (C) the diagram showing equivalent electric circuit for atmospheric scheme (Rycroft, 2006).

A combined action of plasma convection and pitch-angle diffusion of electrons and protons lead to the formation of plasma pressure distribution in the magnetosphere. As it is known, bulk currents are associated to plasma pressure distribution in the magnetosphere. Divergent of these bulk currents gives a spatial distribution of FACs, i.e. magnetospheric sources of ionospheric current systems. Field-aligned currents (FACs) connect the magnetosphere and the ionosphere into a uniform electric circuit (see figure 2). The suggested equivalent electric circuit scheme of the interaction may be analyzed for understanding of the mechanism of geomagnetic activity effect on complex nonlinear system of atmospheric processes.

A very interesting mechanism suggested by Troshichev and Janzhura (2004) needs further considering and improving. Authors in the paper (Troshichev, Janzhura, 2004) noted that the solar wind dynamic pressure effect on the cloud layer would be opposite to that of the interplanetary electric field. Thus, now we can note that, probably, there is some connection between processes at the bow shock front region and meteorological processes at the lower atmosphere, because the magnetospheric plasma convection generation is associated with processes at the bow shock front (e.g., see (Ponomarev, Sedykh, 2006 (b))). However, it is a theme of separate paper.